\documentclass[prb,twocolumn]{revtex4-1} 

\usepackage{amsmath,amsfonts,graphicx}  
\usepackage[hyphenbreaks]{breakurl}
\expandafter\def\expandafter\UrlBreaks\expandafter{\UrlBreaks
  \do\a\do\b\do\c\do\d\do\e\do\f\do\g\do\h\do\i\do\j%
  \do\k\do\l\do\m\do\n\do\o\do\p\do\q\do\r\do\s\do\t%
  \do\u\do\v\do\w\do\x\do\y\do\z\do\A\do\B\do\C\do\D%
  \do\E\do\F\do\G\do\H\do\I\do\J\do\K\do\L\do\M\do\N%
  \do\O\do\P\do\Q\do\R\do\S\do\T\do\U\do\V\do\W\do\X%
  \do\Y\do\Z\do\*\do\-\do\~\do\'\do\"\do\-}%
\usepackage[colorlinks,citecolor=blue,linkcolor=blue,urlcolor=blue,breaklinks]{hyperref}
  
\begin{document}

\title{The Harvard Science Research Mentoring Program}

\author{Or Graur}
\email{or.graur@cfa.harvard.edu}
\affiliation{Center for Astrophysics \textbar\ Harvard \& Smithsonian, 60 Garden St., Cambridge, MA 02138, USA}
\affiliation{Department of Astrophysics, American Museum of Natural History, New York, NY 10024, USA}
\affiliation{NSF Astronomy \& Astrophysics Postdoctoral Fellow}

\date{\today}

\begin{abstract}
 The last decade has seen a proliferation of mentoring programs that provide high-school students authentic research experiences. Such programs expose students to front-line research, equip them with basic research skills (including coding skills), and introduce them to scientist role models. Mentors in such programs range from undergraduate students to faculty members. Here, I describe the founding and first two years of operation of the Harvard Science Research Mentoring Program (SRMP). This program specifically recruits advanced graduate students and postdoctoral scholars to serve as mentors. By mentoring high-school students over a long timescale (September to May), early-career scientists gain hands-on experience in the skills required to advise students---skills that are often required of them in future academic positions yet seldom taught by academic institutions. Finally, I invite directors of existing and prospective SRMPs to join the Global SPHERE Network, through which directors of SRMPs around the world can share their experiences, best practices, and questions.
\end{abstract}

\maketitle

\section{Introduction}
\label{sec:intro}

The number of American workers employed in science, technology, engineering, and math (STEM) occupations has risen rapidly over the last two decades, yet there is an ongoing debate whether this number is sufficient to answer the market's demand.\cite{Noonan2017a,storm1,storm2} Moreover, STEM occupations are far from equitable; women and people of color are significantly underrepresented in STEM jobs relative to their fractions of the total workforce.\cite{Landivar2013,Noonan2017b} This disparity is also seen in the gender and racial makeup of Bachelor's degree holders in physics, as well as that of high-school students taking advanced-placement physics courses.\cite{White2011,Merner2014,Merner2015,Merner2017} 

One way to encourage high-school students to major in STEM fields in college and to achieve equity is to provide them with authentic research experiences along with mentoring by potential role models.\cite{mentor3,mentor2} Such programs have been available for undergraduate students for a while (e.g., the National Science Foundation's (NSF) Research Experiences for Undergraduates\cite{REU}) and have been generally successful.\cite{Seymour2004} Programs that offer such experiences for high-school students, which in this paper will generally be termed ``Science Research Mentoring Programs'' (SRMPs), have been around for several years.\cite{Sousa-Silva2018} 

SRMPs usually pair students with academic advisors (from undergraduate students all the way up to faculty) who supervise them in independent research projects. SRMPs provide students with several benefits: (1) they expose them to modern scientific research; (2) teach them the scientific method; (3) teach them how to think algorithmically and use code to analyze data; (4) sow the seeds of a professional network; and perhaps most importantly, (5) provide them with role models. Altogether, the goal of SRMP is to show students that science and research are not inaccessible ivory towers; that they, too, can become scientists. 

While high-school students are the main target demographic of SRMP, the mentors who work with them also benefit from the program by way of professional development. Although supervising undergraduate and graduate students is an integral part of many scientists' careers, we are not usually trained to do so. SRMP provides graduate students and postdoctoral scholars with hands-on experience in the skills necessary to successfully advise students, such as crafting a project, supervising students' work, and making sure they obtain results by a given deadline. Many mentors stay in touch with their students for years and write them letters of reference (e.g., for college, undergraduate research experiences, or graduate school), thus gaining an additional skill. Finally, participating as a mentor in a SRMP raises the chances of making it onto short lists for faculty positions at academic institutions (such as liberal-arts colleges) that value advising---and mentoring---undergraduate students in short-term, independent research projects.

I served as a SRMP mentor at the American Museum of Natural History (AMNH) between 2011--2016,\cite{AMNH-SRMP} during which time I mentored 17 students. The experience drove me to create my own program at the Center for Astrophysics \textbar\ Harvard \& Smithsonian (CfA) when I moved there in 2016. In this paper, I describe how I set up Harvard SRMP\cite{Harvard-SRMP} and the first two years of its operation, with the hope that my experience founding and directing a SRMP will provide motivation and suggestions for others seeking to found SRMPs at their own institutions. 

\section{SRMP Overview and Timeline}
\label{sec:program}

Harvard SRMP has now completed two years of full operation and is gearing up for its third. The program runs throughout the school year, i.e., from the first week of September to the last week of May. The program begins with an orientation session for students and their parents, during which I describe the program, introduce the mentors and their projects, collect paperwork, administer the first stage of the evaluation survey (see Section~\ref{sec:testimonials}), and take questions.

Throughout September, I meet with the students twice a week, each time for two hours, to administer a very general introduction to astrophysics. In a series of presentations, I introduce basic concepts, from stars and galaxies to imaging and spectroscopy. The goal is for the students to have these concepts in the backs of their minds when they start working on their projects. The students also learn how to use their laptops and the Linux command line. Finally, the mentors give brief introductions to their science and the projects the students will work on. 

At the end of this month, the students rank the projects they would like to work on. Based on this ranking, my knowledge of the students' analytic and coding skills, and the requirements of the projects, I assign the students to their mentors. I also take into account issues of diversity and the importance of mentors serving as role models by asking mentors if they would prefer to work with students who match their gender or ethnicity. To date, one female mentor expressed such a wish and was assigned two female students who had requested her as their first choice. During the first year of the program, 7/10 students received their top pick, two received their second pick, and one their third. In the second cohort, 8/11 students received their top choice and the rest - their second.

Projects should be limited in scope; a task that would take a graduate student or postdoctoral scholar 2--3 weeks to complete will take a high-school student several months. I suggest mentors carve out a small portion of their research program, preferably a project that is not time-critical and that is not required for the success of their overall program. Projects have included: classifying a luminous supernova using photometry and spectra; simulating the dynamics of binaries around super-massive black holes; searching for high-velocity stars in the Milky Way; studying the transits of cometary bodies around a white dwarf; searching for debris disks around binary stars; spectroscopically probing the disk of Epsilon Aurigae; and developing a neural-network approach to measuring galaxy redshifts.

From October to April, the students meet with their mentors twice a week at the CfA, each time for two hours. Because the students are minors, they meet with their mentors in public spaces in the CfA, such as hallways and the CfA's Wolbach Library. Each group of students and mentor comes up with its own schedule, which takes into account the students' other extra-curricular activities and the mentor's research schedule. The schedules are meant to be flexible; some weeks the students will meet their mentor only once, or not at all, usually during the lead-up to midterms and finals, or when the mentor is traveling. The goal is for students to meet with their mentor roughly 100 hours during the program. This forces the mentors to keep an eye on their students' progress and make sure that by the end of the program they have results to present - an important skill for future advisors. 

Many students are so excited by their projects that they choose to continue to work on them from home. As a rule, though, this is not required by the program, and I dissuade mentors from assigning homework.

Most undergraduate research programs are condensed to approximately 10 weeks or less over the summer. The months-long timescale of SRMP, on the other hand, allows students to get stuck, whether because of practical difficulties with executing a particular task or because they need to figure out the next step in the analysis. For most students, this is the first time they are ever required to solve such problems completely on their own, and it teaches them patience and perseverance, two of the more important skills required for actual research. 

Mentors start off by introducing the theory behind the project through assigned readings and discussions during the weekly meetings. This is then followed by hands-on introductions to the tools and data necessary for the project. Finally, the students move on to the analysis itself. 

I encourage mentors to develop their own way of supervising the students. Some mentors are very hands-on, working together with their students, while others are more hands-off, allowing their students to figure things out on their own. As mentors discover throughout the program, the degree to which they engage directly with their students depends on a combination of the students' and mentor's personalities, the students' coding experience (in some instances, mentors spend the first few weeks of the program giving their students a crash-course in coding), and the details of the project. Two of the projects I supervised at AMNH show how a mentor's approach can change from one year to the next. During one year, I let students search for supernovae in \textit{Hubble Space Telescope} data from the CLASH collaboration,\cite{Postman2012} in order to compare the detection efficiency of amateur and professional astronomers, and to check that we had not missed any candidates. In another year, I asked students to measure changes in the periodicity of variable stars in the Magellanic Clouds. The first project was straightforward, and the students searched the data on their own, with minimal supervision. The second project, though, proved to be more difficult, and I worked together with the students through problems with the datasets as well as the analysis techniques. 

Mentors are encouraged to let the students work together as a group in order to introduce them to the principles of collaborative work. I have found that this usually works well in groups of 2--3; larger groups tend to develop power dynamics between the students in which one or two students lead the work and the others are sidelined. When assigning students to mentors, I try to avoid such power dynamics by balancing the personalities of the students. This is hard to do at the beginning of the program, and am helped in this task by my school liaison, who usually has previous knowledge of the students (see Section~\ref{subsec:schools} for more details).

In order to encourage students to conduct outreach activities of their own, they each receive a Galileoscope and tripod, which they use at a star party at their high school. Galileoscopes are cheap but relatively powerful telescopes perfect for observations in cities.\cite{galileoscope} On clear nights, users can see Jupiter's moons and bands, as well as Saturn's rings. For most first-time observers, the Moon, with its craters and dark maria, is just as exciting.

During the last month of the program, the students write up their results in a poster and a short ten-minute talk. These are then presented at a symposium at the CfA to which the students' families are invited. The symposium is also open to CfA researchers, so that the students will be asked questions by their professional peers. I also invite dignitaries from the high school, city hall, and any funding agency that supported the program during that year. The symposium is broadcast live via YouTube and curated by the CfA's Wolbach Library using the Open Science Framework. The posters and talks from the first two symposia can now be viewed online.\cite{symposium1,symposium2}

Throughout the program, I meet with each student and mentor at least twice to learn about their progress in the program and address any problems that might come up. Once a month, I meet with the entire cohort to go over a specific topic or skill. Topics have included a description of the academic ladder, gender and racial biases in academia, and how to write and present a poster and a talk. For the latter, I prepare templates that the students then fill out and personalize.

\section{Setting Up SRMP}
\label{sec:howto}

Below, I outline the three steps I took to set up my SRMP in the span of a single academic year: (1) partnering with a local school, (2) recruiting students and mentors, and (3) securing funding. Before setting up your own program, I suggest consulting online outreach resources for information and advice. For astrophysics programs, for example, there is the Menu of Outreach Opportunities for Science Education, provided by the American Astronomical Society (AAS).\cite{moose}

Recently, several SRMPs, including Harvard SRMP, have banded together to create the Global SPHERE Network,\cite{SPHERE} a website that any SRMP serving high-school students is welcome to join. This website serves two functions: (1) to help students find a nearby program; and (2) to allow program directors to share questions and best practices. If you are about to set up your own program, or if you already run a program, please consider joining the Global SPHERE Network and adding your experiences to the mix.

\subsection{Partnering with local schools}
\label{subsec:schools}

Many academic institutions will have an education/outreach office or officer who may already have connections at the local high schools. I strongly advise checking for such existing connections before attempting to foster your own, as finding inroads into local high schools can be the most time-intensive step of setting up a SRMP. I also suggest checking the schools department at your local city hall. Some, such as Cambridge, will have a person in charge of STEM development---a natural point of contact for a SRMP. 

Through the education research department at the CfA, I was put in contact with a student at Cambridge Rindge and Latin School (Cambridge's sole public high school, alongside two additional, charter high schools) who was attempting to create an aerospace engineering and astronomy club at the school. I wrote the student a letter of support, which aided his effort. Later on, I recruited postdoctoral scholars and graduate students to lead hands-on astrophysics activities at this club. Each activity was spread over two meetings. Instead of showing up and giving a frontal lecture, the speaker prepared ahead of time a syllabus that was shared with the students and allowed them to learn the necessary introductory material on their own. The speaker would then show up for the second meeting, provide a short recap of the introduction, and then proceed to devote most of the meeting ($\sim 45$ minutes) to directing the activity. Activities have included creating color images from \textit{Hubble Space Telescope} data, simulating a supernova light curve using an Arduino computer, and observing the Moon with the 9-inch Clark Telescope on the roof of the CfA. These syllabi are available from the Harvard SRMP website and free to use.\cite{lectures}

Through the hands-on lecture series, I got to know the school's astronomy teacher, Mr. Tal SebellShavit, who then became my main point-of-contact at the school. Without his help I would not have been able to set up my SRMP as quickly as I did. He organized a schedule for my recruitment presentations, shared his knowledge of the candidates, organized their interviews, and helped the selected students prepare the paperwork for their stipends (see Section~\ref{subsec:funding}, below).

\subsection{Recruiting students and mentors}
\label{subsec:recruit}

\subsubsection{Students}

Each semester, I advertise the program at the school by giving short, ten-minute presentations in every science class throughout the day. Beginning with the second year of the program,  alumni from the previous year's program tag along and describe their projects and experiences. This makes it easier for interested students to ask questions and see themselves as potential candidates.

At the end of each presentation, I hand out application packets and leave a few behind for those students too shy to take them directly from me. The packet, available on the Harvard SRMP website,\cite{Harvard-SRMP} includes a standard application form asking for the student's name, address, etc., as well as instructions for completing two essays.

Several studies have shown that letters of recommendation tend to reflect the gender biases of their writers.\cite{Dutt2016} Students' grades are not immune from bias either, whether it stems from their teachers\cite{Riegle2012,Lavy2015} or the students themselves (e.g., through ``stereotype threat'').\cite{Steele1995,Steele2002} 

Instead of transcripts and letters, I ask students to submit two short essays, up to one page each. The first is a personal essay in which they introduce themselves and explain why they are interested in science and astrophysics, why they want to join the program, whether they have any previous research experiences (this is not a prerequisite), and whether they have any special hobby or talent they would like to share. The goal of this essay is to gauge the students' interest in the program along with their writing and self-expression skills. 

For the second essay, the students are asked to choose an image from the website Astronomy Picture of the Day.\cite{apod} This website publishes a daily astronomy image along with a short, one-paragraph description. The students are asked to expand on this blurb. The goal here is to get a feeling for the students' ability to engage with a topic they have never seen before and learn about it on their own. For both essays, I provide a list of leading questions to guide the students through these tasks.

The essays are read by myself and by the program's liaison at the high school, who either already knows the students or can ask other teachers about them. I use the essays to get a first impression of the students and then follow up by interviewing them at the high school. The interviews are short (typically 5--10 minutes) and are meant to learn more about the students' prior research experiences, coding skills, and personality. 

With this information, I select the students according to the following criteria, in this order: (1) Seriousness and excitement, as apparent from the application essays and interview; (2) prior research and coding experience, with the aim of reaching both experienced and inexperienced students; (3) age, where seniors are prioritized over juniors; (4) gender, with the aim of achieving gender parity in each cohort; and (5) ethnicity, with the aim of having an ethnically-diverse cohort. Depending on the mix of candidates in each given year, some skilled, experienced students may be turned away if they are juniors who would be able to apply for the program again in the future. This has happened once, and the student in question reapplied and was accepted into the second cohort. No student is either accepted or turned away solely based on their gender and ethnicity. I explain this list of priorities in my recruitment presentations so that applicants are aware of the basis for my decisions.

Over the first three application cycles, 50--100 students took application packets and 15, 13, and 26 students, respectively, submitted completed applications. During recruitment for the 2018--2019 cohort, four applications were received from students who do not study in Cambridge, including out-of-state students, raising the total applicant pool to 17. Of these, I selected one student, who runs her own outreach program for girls in elementary schools, to form a cohort of 11 students. Similarly, the 2019--2020 recruitment cycle included two applications from students outside Cambridge, raising the total to 28. The 100\% jump in applications between years 2 and 3 may be construed as a growing interest in the program. This will be tested by application statistics in future cycles.

The Cambridge, MA school district serves a diverse student population that is not significantly dissimilar from other populations across the nation. Cambridge Rindge and Latin School enrolls $\sim 2000$ students from across a socio-economically diverse school district. In 2011--2012, the district's enrollment by race/ethnicity was: 31.4\% African-American, 11.1\% Asian, 13.5\% Hispanic, 0.6\% Native American, 38.5\% White, 0.3\% Native Hawaiian / Pacific Islander, and 4.6\% non-Hispanic multi-race. Students from low-income families accounted for 42\% of the district, special-education recipients made up 21\% of the district, and 7\% were English-language learners.\cite{CRLSreport}

Using the same race/ethnicity division as above, the first two Harvard SRMP cohorts were: 5\% African-American, 25\% Asian, 10\% Hispanic, 0\% Native American, 45\% White, 0\% Native Hawaiian / Pacific Islander, and 15\% non-Hispanic multi-race. Five students (25\%) were eligible for free/reduced lunches, one student (5\%) qualified for special education, and five students (25\%) came from homes were English was a second language.

With the caveat that the above analysis suffers from small-number statistics due to the current sample of 20 students (the student who does not live in Cambridge is not included in this calculation), it is clear that the selection criteria described above result in diverse cohorts that overall reflect the ethnic/racial and socio-economic diversity of the Cambridge, MA school district. However, there is also clearly room for improvement, such as in recruiting more African-American students to the program. To increase the diversity of the applicant pool, during the third application cycle I emphasized the diversity of previous cohorts, encouraged students to put aside their conceptions of what scientists look like, and had several alumni tag along and describe their projects. Additionally, my liaison at the school periodically asked teachers to remind students to submit applications. These steps may have had the desired effect, as the breakdown of the 26 applicants from Cambridge was 23\% African-American, 34\% Asian, 31\% White, and 12\% non-Hispanic multi-race.

\subsubsection{Mentors}

Mentors can be anyone from undergraduate students to emeriti faculty members. I prefer to look for mentors among the advanced graduate students and postdoctoral scholars at my institution, as they would benefit the most from the experience. Sadly, although many academic careers involve advising students, the necessary skills are seldom taught in graduate school. SRMPs allow early-career scientists to acquire these skills through hands-on experience with students in a non-threatening environment (i.e., their future hiring or tenure decisions are not dependent on the success of their students).

Graduate students should have completed their course requirements, passed their qualifying exams, and be well on their way to completing their research projects. At this stage, they have enough knowledge and expertise to pass on to other students and can begin to think of small research projects to spin off of their main Ph.D. project. For graduate students, the SRMP experience can help their search for a postdoctoral position, as some fellowships and grants require applicants to devise education and public outreach (EPO) programs or explain the broader impacts of their work.

For postdoctoral scholars, SRMP is not only an opportunity to learn how to become an advisor but also a way to try out new subject fields or projects that would otherwise be too risky or time consuming. It is also a way to signal to potential employers that they will already know how to work with students and successfully lead them to complete their degrees. For those seeking positions at liberal arts colleges, it is important to note that working with high-school students is akin to working with undergraduate students and that the scope of the research projects assigned to them is often similar.

I recruit five mentors each year, so that each mentor is assigned two students. As in other institutions, the mentors are funded by a mix of independent fellowships and grants. Postdocs with independent fellowships manage their time as they see fit. By directing mentors to assign their students small portions of their own research programs, I also make it possible for postdocs and graduate students funded by specific grants to comply with the grants' requirements on their time.

I aim for a diverse cohort of mentors but have found it hard to recruit female mentors, with only one candidate stepping forward each year. I see four possible reasons for this failing: (1) women in academia receive more invitations than men to provide academic services such as serving on committees and in programs such as SRMP,\cite{Guarino2017} thus placing a heavier burden on their time; (2) the fraction of female postdocs and graduate students at the CfA, as in most departments, is still smaller than the fraction of men; (3) I have yet to secure an incentivizing financial award for mentors; and (4) there is a (mis)conception among postdocs and graduate students, especially in research-heavy institutions, that promotion up the academic ladder depends solely on one's research output.\cite{Graur2018} The last point affects male academics as well, which is why, to date, each year only five candidates in total have answered my call. However, two of the first year's mentors chose to return for a second year.

\subsection{Funding}
\label{subsec:funding}

The main funding requirement for a SRMP is offsetting the time spent on it by the program's director. In my case, I spend $\approx 0.1$--$0.2$ FTE on the program, which is made possible by a NSF Astronomy and Astrophysics Postdoctoral Fellowship. Such funding can also be acquired through the Broader Impacts section of general NSF grant applications or by applying to private foundations, such as the Sloan or Simons Foundations. Recently, the Heising-Simons Foundation created a new prize postdoctoral fellowship, the 51 Pegasi b Fellowship in Planetary Astronomy,\cite{Pegasi} which requires candidates to pursue activities that will advance diversity, equity and inclusion, a goal shared by many SRMPs. Social clubs, such as the local branch of Rotary International,\cite{rotary} can also be approached for small grants. Professional associations, such as the AAS or the American Physical Society, also provide small grants for EPO programs.\cite{AAS,APS} Finally, I suggest consulting with your institution's public affairs, communications, or development departments, who might be able to pitch SRMP as a way to strengthen the institution's ``town-and-gown'' relationship. Press releases and news articles generated by the program are a useful way to catch these departments' attention.\cite{press,gazette}

Aside from the director's salary, I suggest securing additional funding for stipends and laptops for the students. This removes two major barriers to participation in the program, as not all students may be able to afford a computer, and some students, especially those from underserved communities, may have to choose between participating in the program or finding work after school hours to supplement their families' incomes. 

\textit{Stipends}: If students spend four hours a week on their projects throughout the school year, that usually comes out to a rough total of $\approx100$ hours. Based on a minimum wage of \$10--15 an hour, I recommend stipends of \$1,000--1,500 per student. Besides private foundations, local government can be a good source for the necessary funds. I approached a Cambridge City Councillor (Nadeem Mazen), whose agenda included education, who then connected me to the right people in city hall. For the last three years, the City of Cambridge has provided \$15,000 a year for stipends. Cambridge Rotary has provided a stipend for the 11th student in the 2018--2019 cohort, who does not live in Cambridge.

\textit{Computers}: Although desktop computers are cheaper, on average, than laptops, I suggest procuring the latter, so they can be disbursed to the mentors for their students' use each year. Many academic institutions have contracts with specific computer companies that allow the purchase or rental of relatively powerful laptops for less than \$1,000 each. Alternatively, some computer and hi-tech companies have been known to donate computers to schools and EPO programs.\cite{IBM}

\section{Evaluation and Testimonials}
\label{sec:testimonials}

Of the 21 students who took part in the first two cycles of the program, 19 remained highly engaged with their projects throughout the year. One student began to lose interest towards the end of the first year, and another student---due to personal issues unrelated to the program---quit several weeks before the second year's symposium. Student testimonials, some of which are quoted below, show that all students, including the student who ended up quitting the program, were enriched by the SRMP experience.

In order to evaluate what impact the program may have had on the students, the City of Cambridge devised a short survey that I administer to the students on the first and last days of the program. Among other things, the survey asks the students how they feel about their comeptence in, e.g., math and coding; if they know adults who work in various scientific fields; and how interested they are to pursue ``traditional'' STEM-related careers, such as research, coding, and engineering careers, as well as careers in non-STEM fields, such as education, government, and journalism. Several more cohorts of students are required to construct a sample large enough for a statistically meaningful evaluation, but so far perliminary results echo the testimonials recorded below. 

In the meantime, the program has already had a couple of tangible results: five peer-reviewed papers are in preparation (two papers by Ginsburg et al., \textit{in prep.}; Ravi et al., \textit{in prep.}; Villar et al., \textit{in prep.}; and Zhou et al., \textit{in prep.}), and one student has received an internship at a computational biology startup that his mentor had joined. Other SRMPs, such as ORBYTS\cite{Sousa-Silva2018}, the Science Internship Program at University of California, Santa Cruz,\cite{SIP} and AMNH SRMP, have also produced peer-reviewed publications.\cite{Tollerud2012,Yang2013,Kirby2013,Kirby2015,Kirby2016,Toloba2016,Prichard2017,McKemmish2017,Chubb2018,Graur2014,Leigh2016,Leigh2017,Leigh2018,Ibragimov2018} More broadly, it is now abundantly clear that high-school students, given the right projects, can produce publishable results. A search of the SAO/NASA Astrophysics Data System for peer-reviewed papers published since 2000 with co-authors affiliated with high schools found 176 papers published in various astrophysics journals, 60 papers published in the different Physics Review journals, and 7 papers published in Nature and Science.

Testimonials from the first cohort indicate that coding skills are a major skill gained through the program, e.g., \textit{``While I have experience with coding in Java, I did not know python before this program,'' ``For me, almost everything we have done has been completely new --- I had to learn Python from almost never having used it before,''} and \textit{``I have never coded before this project, so coding in python has been entirely new to me.''} 

Students also enjoyed working on real problems and contributing to their mentors' research programs, e.g.: \textit{``I am excited to go meet with my mentor every week and enjoy the work itself. I usually spend parts of my free time working on my current project. It feels great to know that I am contributing to his work and research.''} 

Although the research was hard at times, students found this engaging rather than frustrating: \textit{``I feel that the topic that I'm working with is hard, but that doesn't mean it's bad, and additionally, you get to work with an expert, making the program both challenging and engaging, which I enjoy.''}

Finally, though it would be hubris to assume that a nine-month program could change the course of a student's life, a SRMP does allow students to ``test-drive'' a science research career. Roughly half of the students in the first two cohorts had already decided in what fields they wanted to major in college before they joined the program (from STEM fields such as physics, astronomy, engineering, and computer science to the arts and humanities, including animation, drama, and political science). For these students, the program either exposed them to astrophysics for the first time, or helped them decide between several fields of interest. One student said: \textit{``I already knew I wanted to major in political science but the most important thing it did for me was to spark interest in the study of the universe.''} Another student, who no intends to major in engineering, noted: \textit{``Although I have always been intrigued by astrophysics, SRMP has definitely given me direction in understanding how computer science is often intrinsic to astronomical research. Because I love both computer science and astronomy, SRMP was a perfect gateway into getting a better sense of what path I want to go down in college.''}

Some students joined the program because they were curious about astronomy and whether they wanted their future careers to include research components. Thus, one student came in wanting to major in social work but is now wondering how to combine practice and research, and is also considering neurology. Another student concluded that, although she would not want to pursue a research career, she would like to find ways to combine science and art. A third student used the program to explore the option of a career in science journalism and eventually decided she would rather pursue research in physics and astronomy. In her own words: \textit{``Coming into the program, I had a vague idea of where I wanted to take my interest in astronomy. After spending the past few months alongside my mentor, I've now gained more insight on what astrophysical research entails, and I will be pursuing such a career next year in college. SRMP has helped me solidify my aspirations, and I am so glad to have been a part of it.''}

The program also contributed to the mentors' professional development. One mentor described his experience in these words: \textit{``Participating in the SRMP provided my first opportunity to act as primary mentor for students engaged in a research project. This was extremely useful for me, as it can be quite difficult to gain experience in teaching/mentoring at a postdoctoral level, despite such skills being essential if one wishes to pursue academia at the faculty level. The fact that the program lasted for the full academic year was important, in that it gave me time to adjust to the role. I found it challenging to set a project that was difficult enough to stimulate the students, without being too difficult. Having a year to complete the project gave me time to find my feet as a mentor and establish the right amount of assistance to give to the students. I particularly liked that the students came from under-represented demographics in STEM research---this gave me a chance to learn inclusive practices, and increased my desire to take on more such students in the future. Along with these actual skills that I gained, being able to add the experience to my CV has already proved helpful in the academic job market.''}

Another mentor noted the importance of the program's long timescale for the students'---and his own---professional development: \textit{``By working with the students for an entire academic year, we were able to work on developing a whole suite of skills, instead of rushing for project success. The projects required some Python scripting, but they mostly relied on the students finding the inquiry for the next step in the research. The students learned to problem solve, and come up with ideas for potential solutions, and I learned to guide them through the problems, and provide help only where needed.''}

Half of the mentors who served in the program's first year chose to return for a second year --- a strong vote of confidence in the program's usefulness to their research program and professional development.

\section{Conclusion}
\label{sec:conclusions}

In this paper, I have described the setup and first two years of the Harvard Science Research Mentoring Program (SRMP), with the hope of motivating and providing suggestions for others interested in starting their own SRMP. 

SRMPs can have a profound impact on both students and mentors. Students gain valuable research skills and the beginning of a professional network. I have personally stayed in touch with some of my students all the way from high school to graduate school and helped them along the way, e.g., with letters of reference and invitations to conferences. Mentors, most of them for the first time, gain hands-on experience supervising students. Successful mentoring requires a valuable set of skills that is expected by most academic positions yet seldom taught. SRMP is one of the only venues through which graduate students and postdoctoral scholars can gain these skills.

Harvard SRMP currently accepts ten students each year. This number was chosen to minimize the time I spend managing the students, as I currently devote only $\sim10$--$20$\% of my time to the program. It is also easier to convince funding agencies to support a program just starting out if the amount of funding request is small, commensurate with the number of students. Once the program strikes roots and is no longer seen as a ``pilot,'' it is natural to think of expansion: to other schools, other subject fields, other towns. The AMNH SRMP encompasses biology, astrophysics, and geophysics. The Science Internship Program at the University of California, Santa Cruz started off with three students in 2009 and has since expanded to include $>150$ students working in 14 departments. I am currently working on expanding Harvard SRMP to MIT and the two charter high schools in Cambridge.

Although I founded my SRMP at Harvard University, SRMPs can be set up at any institution that employs researchers who are enthusiastic about public outreach. The scale of the program and its funding model may change but the main aspects---recruiting high-school students to work on independent research projects alongside early-career scientists---should be applicable across the US. The Global SPHERE Network is based on the assumption that this model is also applicable across the world. Thus, along with the SRMPs at Harvard, AMNH, UC Santa Cruz, and University College London already mentioned above, the Global SPHERE Network also includes SRMPs at Academia Mexicana de las Ciencias; University of Minnesota; Evergreen Valley College; University of Goettingen; Michigan State University; and the University of Nigeria, Nsukka.\cite{partners} The New York City Science Research Consortium comprises 13 programs at various institutions across the city, from AMNH, Columbia, and NYU to the City University of New York, Genspace, the Rockaway Waterfront Alliance, and The Rockefeller University.\cite{consortium}

With the recognition that mentoring is essential to expanding the STEM workforce and to promoting the participation of women and people of color, there is a newfound openness among academic institutions, schools, and funding agencies to support such SRMPs. Mentoring students at AMNH was one of the highlights of my graduate studies, so much so that it drove me to create my own SRMP at the CfA. I hope that this paper will motivate and assist anyone interested in creating a similar program. When you do, come join us on the Global SPHERE Network and share your own experiences.

\begin{acknowledgments}

I thank the anonymous referees for their comments and suggestions. I thank Tal SebellShavit for being my liaison at Cambridge Rindge and Latin School; George Hinds for managing the students' stipends; Sharlene Yang for preparing the program's evaluation survey; Susan Walsh and Susan Mintz for shepherding the program's funding from the City of Cambridge; Katie Frey for managing the students' laptops; Eric Brownell for setting up the program's OSF page; and Judith Schwab for handling the program's administrative aspects. Finally, I thank the mentors of the 2017--2019 cohorts: Idan Ginsburg, Jae Hyeon Lee, Matt Nicholl, Vikram Ravi, Sandro Tacchella, Ashley Villar, and George Zhou. This work was supported by an NSF Astronomy and Astrophysics Fellowship under award AST-1602595, as well as by the City of Cambridge, the Wolbach Library at the CfA, Cambridge Rotary, and generous anonymous donors. This research has made use of NASA's Astrophysics Data System and the NASA/IPAC Extragalactic Database (NED) which is operated by the Jet Propulsion Laboratory, California Institute of Technology, under contract with NASA.

\end{acknowledgments}

\end{document}